\documentclass[aps, amssymb, preprint,nofootinbib, 12pt]{revtex4}
\draft

\begin{document}

\parbox{4cm}{\vspace*{2cm}}
\title{Fermionic zeromodes in heterotic fivebrane backgrounds}
\author{Yasutaka Katagiri}
\thanks{e-mail: kata@phys.metro-u.ac.jp}
\address{Department of Physics, Tokyo Metropolitan University, Hachioji, Tokyo 192-0397, Japan}
\date{\today}
\author{Noriaki Kitazawa}
\thanks{e-mail: kitazawa@phys.metro-u.ac.jp}
\address{Department of Physics, Tokyo Metropolitan University, Hachioji, Tokyo 192-0397, Japan}

\begin{abstract}
We investigate the explicit form of the fermionic zeromodes in heterotic fivebrane backgrounds.
By explicitly solving the fermionic field equations in fivebrane backgrounds, two normalizable and physical fermionic zeromodes are obtained.
Each of these zeromodes has a non-vanishing gravitino component.
We suggest a possible scenario of the gravitino pair condensation.
\end{abstract}

\maketitle

\newpage

\section{Introduction}
Supersymmetry is a very attractive symmetry, but we do not have any concrete experimental evidences which support this symmetry at low energies.
Therefore it must be broken spontaneously in some way.
It is desirable that supersymmetry is broken dynamically, and one interesting scenario was proposed in Ref.\cite{Witten}.
In this scenario, local supersymmetry is broken due to the topological effect of the gravity.
As an explicit realization of this scenario, four-dimensional supersymmetry breaking in the background of Eguchi-Hanson metric is well-known\cite{Konishi}.
In this special background there are non-trivial gravitino zeromodes, and these zeromodes induce the gravitino condensation $\langle \psi_{ab}\psi^{ab} \rangle \ne 0$ which means the breaking of supersymmetry through the Konishi anomaly relation.

It is an interesting question whether this scenario is possible in the string theory, because the string theory can be considered as a consistent theory of the quantum gravity.
The heterotic fivebranes, or the Callan-Harvey-Strominger (CHS) solitons, are known as exact solutions of the heterotic string theory\cite{CHS}.
These solitons keep just a half of the supersymmetry of the heterotic string theory.
In Ref.\cite{Rey} a background field configuration, which is similar to the one of the heterotic fivebrane, was studied, and the existence of the gravitino zeromodes which can contribute to the gravitino condensation was explicitly shown.
Therefore, we can expect the existence of gravitino zeromodes also in the heterotic fivebrane background.

The supersymmetry breaking in ten-dimensional spacetime of the heterotic string theory has no direct relation to the supersymmetry breaking in the real world.
We need further ideas to construct realistic models.
But it is interesting to study the dynamics of fermions in the string theory which could trigger some interesting phenomena like the supersymmetry breaking as a first step.

In this paper, we find the explicit form of two fermionic zeromodes in the heterotic fivebrane background by explicitly solving the field equations of fermions.
They can be considered as well-defined physical zeromodes which can contribute to the gravitino condensation.

This paper is organized as follows.
A short review of the heterotic fivebrane is given in the next section.
In Sec.\ref{sec:three}, we explicitly derive the field equations for fermionic zeromodes.
In Sec.\ref{sec:four} we give the explicit solution of these field equations.
The argument of physical conditions of these fermionic zeromodes is also given in this section.
In Sec.\ref{sec:five} a possible scenario of the gravitino pair condensation is suggested.

\section{A short review of the heterotic fivebrane}
\label{sec:two}
The soliton is the solution of the classical field equation, and it is distinguished from other vacuum solutions by being translationally non-invariant.
In the string theory, many objects with solitonic properties have been discovered.
Especially, the heterotic fivebrane, or the CHS soliton, is one of the important solitons.
It is an exact solution of the heterotic supergravity, and moreover, its validity as a solution of the heterotic string theory can be confirmed using the conformal field theory in a certain limit.
In this section, we give a short review of the construction of this soliton.

The action of the heterotic supergravity is given in Ref.\cite{BdR}.
After the field redefinition of Ref.\cite{Bellisai},
\begin{eqnarray} %%% Field redefinition of Bellisai %%%
H^{BdR} = \frac{\sqrt{2}}{3}H,\:\:\:
\lambda^{BdR} = \sqrt{2} \lambda,\:\:\:
\phi^{BdR} = e^{\frac{2}{3} \Phi},
\end{eqnarray}
it becomes
\begin{eqnarray}  %%% heterotic SUGRA action %%%
S_{hetero} &=& \int d^{10}x \left\{ {\cal L}_{B} + {\cal L}_{F} \right \}, \label{S_Hetero}\\
{\cal L}_{B} &=& %%% Bosonic part %%%
\sqrt{g} e^{-2\Phi} \left\{
-\frac{1}{2} R(\omega)
-\frac{1}{6} H_{MNP} H^{MNP}
+2\partial_M \Phi\partial^M \Phi
-\frac{1}{4} F^{\alpha}_{MN} F^{\alpha MN}
\right\}, \label{bosonic_Lagrangian} \\
{\cal L}_{F} &=& %%% Fermionic Part %%%
\sqrt{g} e^{-2\Phi}
\left\{
-\frac{1}{2} \overline{\psi}_M \Gamma^{MNP} D(\omega)_N \psi_P
+8\overline{\lambda} \Gamma^M D(\omega)_M \lambda
+4\overline{\lambda} \Gamma^{MN} D(\omega)_M \psi_N
 \right. \nonumber \\
&&{}\left.
+4\overline{\psi}_M\Gamma^N\Gamma^M\lambda\partial_N\Phi
-\overline{\psi}_M\Gamma^M\psi_N\partial^N\Phi
-\frac{1}{2}\overline{\chi}^\alpha\Gamma^M{\cal D}(\omega, A)_M\chi^\alpha
\right.\nonumber \\
&&{}\left.
+\frac{1}{24} H^{RST} \left[
\overline{\psi}_M \Gamma^{[M} \Gamma_{RST} \Gamma^{N]} \psi_N
+8\overline{\psi}_M {\Gamma^{M}}_{RST} \lambda
-16\overline{\lambda} \Gamma_{RST} \lambda
\right]
\right. \nonumber \\
&&{}\left.
-\frac{1}{4} \overline{\chi}^\alpha \Gamma^M \Gamma^{NR} F^\alpha_{NR}
\left(
\psi_M
+\frac{2}{3} \Gamma_M \lambda
\right)
+\frac{1}{24} tr\overline{\chi} \Gamma^{MNR} \chi H_{MNR}
\right\}, \label{fermionic_Lagrangian}
\end{eqnarray}
where $M, N, \ldots$ are indices of ten-dimensional spacetime coordinates, $\alpha$ stands for $E_8\times E_8$ adjoint index, and $\Gamma^{MNP\ldots}$ are totally anti-symmetrized gamma matrices.
They are normalized as $\Gamma^{MN} = 1/2!(\Gamma^M \Gamma^N -\Gamma^N \Gamma^M)$.
This action has ${\cal N}=1$ local supersymmetry.
The three-form tensor field $H$ plays an important role in the construction of the heterotic fivebrane.
It is related to the two form potential $B$ as described in Ref.\cite{GS}.
\begin{eqnarray}  %%% Definition of H field %%%
H = dB 
+\alpha' \left(\omega_{3}^{Lorentz}(\Omega_{+})
-\frac{1}{30} \omega_{3}^{YM}(A) \right) +\ldots, \label{H_definition}
\end{eqnarray}
where dots mean higher-order $\alpha'$ corrections and $\omega_{3}^{Lorentz} (\Omega_{+})$ and $\omega_{3}^{YM} (A)$ are the Chern-Simons three-forms whose exterior derivatives give $Tr R(\Omega_{+}) \wedge R(\Omega_{+})$ and $tr F(A) \wedge F(A)$, respectively.
Taking the exterior derivative of Eq.(\ref{H_definition}), the famous anomalous Bianchi identity is obtained.
\begin{eqnarray}   %%% Anomalous Bianchi Identity %%%
dH = \alpha' \left(Tr R(\Omega_{+}) \wedge R(\Omega_{+})
-\frac{1}{30} tr F(A) \wedge F(A) \right) + \ldots. \label{Anomalous_Bid}
\end{eqnarray}
We define the generalized spin connection $\Omega_\pm$ as
\begin{eqnarray}   %%% Definition of \Omega %%%
{\Omega_{\pm M}}^{AB} = {\omega_M}^{AB} \pm {H_M}^{AB}, \label{definition_Omega}
\end{eqnarray}
where ${\omega_M}^{AB}$ is the usual spin connection (the sign of our $\omega$ is opposite in comparison with the one in Ref.\cite{BdR}), and $H$ plays a role of the tortion in $\Omega_\pm$.
In order to obtain the solitonic solution, it is convenient to look for the bosonic backgrounds which are annihilated by a part of supersymmetry transformations.
Only the vacuum is annihilated by all of the supersymmetry variations.

In zero fermionic backgrounds, supersymmetry variations take especially simple form.
All the variations of bosonic fields vanish, and the variations of fermionic fields are\cite{BdR}
\begin{eqnarray}  %%% Fermionic part of SUGRA variation %%%
\delta\psi_M &=&
\left(\partial_M + \frac{1}{4} {\Omega_{-M}}^{AB} \Gamma_{AB} \right)\epsilon, \label{GrV}\\
\delta\chi^\alpha &=& 
-\frac{1}{4} F_{MN}^{\alpha} \Gamma^{MN} \epsilon, \label{GaV}\\
\delta\lambda  &=& 
-\frac{1}{4} \left(\Gamma^M \partial_M \Phi -\frac{1}{6} H_{MNP} \Gamma^{MNP} \right) \epsilon, \label{DlV}
\end{eqnarray}
where the infinitesimal parameter $\epsilon$ belongs to the {\bf 16} dimensional Majorana-Weyl spinor representation with positive ten-dimensional chirality.

Now we consider the solitonic solution with the fivebrane structure.
We assume that all the fields are independent of the coordinates along the fivebrane.
Then, the Lorentz symmetry is decomposed into $SO(5,1) \times SO(4) \subset SO(9,1)$, and the supersymmetry variations are parameterized by the spinors of $\epsilon_{+} \oplus \epsilon_{-} = (4,2_{+}) \oplus (4^*,2_{-})$.
It can be verified that the following backgrounds are annihilated by just a half of the supersymmetry transformation parameterized by $\epsilon_+$.
\begin{eqnarray}   %%% Heterotic Fivebrane solution %%%
F_{\mu\nu} &=& \tilde{F}_{\mu\nu}, \\
H_{\mu\nu\rho} &=& -\sqrt{g} \epsilon_{\mu\nu\rho\sigma} \partial^\sigma \Phi, \label{H} \\
g_{\mu\nu} &=& e^{2\Phi} \delta_{\mu\nu}. \label{metric}
\end{eqnarray}
Here, indices $\mu, \nu, \ldots$ stand for the four-dimensional Euclidean coordinates transverse to the fivebrane.

The explicit form of the dilaton field $\Phi$ is not determined yet.
It can be determined using the anomalous Bianchi identity of Eq.(\ref{Anomalous_Bid}).
With Eq.(\ref{H}), this identity can be rewritten as
\begin{eqnarray}   %%% Anomalous Bianchi Identity in the heterotic fivebrane %%%
e^{-2\Phi} \square e^{2\Phi} =
\alpha' \left(Tr R(\Omega_{+}) \wedge R(\Omega_{+})
-\frac{1}{30} tr F(A) \wedge F(A)\right) +\ldots. \label{Anomalous_Bid_In_CHS}
\end{eqnarray}
Here, we take the gauge field configuration as $SU(2)$ selfdual instanton embedded in the $SU(2)$ part of $E_8 \supset SU(2) \times E_7$, and make the following identification.
\begin{eqnarray}   %%% Standard embedding %%%
R_{\mu\nu} (\Omega_{+})^{ab} = \frac{1}{2} \overline{\eta}^{Iab} F^I_{\mu\nu}, \label{Std_Embedding}
\end{eqnarray}
where $I$ is the $SU(2)$ adjoint index and $\overline{\eta}$ is the 't Hooft's anti-selfdual $\eta$  symbol.
To realize this identification, $R_{\mu\nu} (\Omega_{+})$ should be anti-selfdual.
The anti-selfduality is ensured by the condition of $\square e^{2\Phi} = 0$, where $\square$ is the usual four-dimensional Laplacian $\delta^{ab} \partial_a \partial_b$.
As a result of this identification, the right-hand side of Eq.(\ref{Anomalous_Bid_In_CHS}) vanishes, and we obtain an exact formula: $e^{-2\Phi} \square e^{2\Phi} = 0$.
This procedure is the analogy of the familiar trick used in the Calabi-Yau compactification of the heterotic string theory (standard embedding, see Refs.\cite{Calabi-Yau_1, Calabi-Yau_2}).
As a solution of the equation $\square e^{2\Phi} = 0$, we adopt the following dilaton field $\Phi$ which was given originally in Ref.\cite{CHS} (other type of solutions were found in Refs.\cite{Dilaton_Sol_1, Dilaton_Sol_2, Dilaton_Sol_3}):
\begin{eqnarray}   %%% Dilaton configuration %%%
e^{2\Phi} = e^{2\Phi_0} + \sum_I \frac{{\cal Q}_I}{\left( x - x_{0I} \right)^2},
\qquad {\cal Q}_I \in {\bf Z}\label{Dilaton},
\end{eqnarray}
where the constant term $e^{2\Phi_0}$ is determined by the value of the dilaton field at the infinity and $x_{0I}$ is the position where the Ith fivebrane is located.
${\cal Q}_I$ are interpreted as the charges of heterotic fivebranes, and they are quantized as the integral multiple of $\alpha'$\cite{Charge_Quantization}.
The global geometry of this soliton is considered as half-cylinders glued into a flat Euclidean space, or a collection of semi-wormholes.
We will consider only the single fivebrane from now on.

\section{Field Equations for Fermionic Zeromodes}
\label{sec:three}
In the previous section, the concrete structure of the heterotic fivebrane was introduced.
We are now ready to discuss about the fermionic zeromodes in this backgrounds.
Following Ref.\cite{Bellisai}, fermionic field equations are explicitly derived from the action of Eq.(\ref{S_Hetero}).
\begin{eqnarray}   %%% 10-dimensional Fermionic equations %%%
e^{-2\Phi}&&\bigg[
\Gamma^{MNR} (D_N - \partial_N\Phi) \psi_R
-2\partial^{[M} \Phi \Gamma^{N]} \psi_N
-\frac{1}{12} H^{RST} \Gamma^{[M} \Gamma_{RST} \Gamma^{N]} \psi_N \nonumber \\
&&+4\Gamma^{MN} (D_N - \frac{3}{2} \partial_N \Phi) \lambda
-6\partial^M \Phi \lambda
-\frac{1}{3} H^{RST} {\Gamma^M}_{RST} \lambda
+\frac{1}{4} F_{RS} \Gamma^{RS} \Gamma^M \lambda
\bigg] = 0, \label{10D_Gravitino_Eq}\\
e^{-2\Phi } && \bigg[
-8\Gamma^M (D_M - \partial_M) \lambda
+\frac{2}{3} H^{RST} \Gamma_{RST} \lambda
-2\Gamma^{MN} (D_M - \frac{3}{2} \partial_M \Phi) \psi_N \nonumber \\
&& -(3\partial^M \Phi - \frac{1}{6} H^{RST} {\Gamma_{RST} }^M) \psi_M
-\frac{1}{24} F_{NR} \Gamma_M \Gamma^{NR} \Gamma^M \chi
\bigg] = 0, \label{10D_Dilatino_Eq}\\ 
e^{-2\Phi} && \bigg[
\Gamma^M ({\cal D}_M - \frac{3}{2} \partial_M) \chi
-\frac{1}{12} H_{MNR} \Gamma^{MNR} \chi
+\frac{1}{4} \Gamma^M \Gamma^{NR} F_{NR} (\psi_M + \frac{2}{3} \Gamma_M) \lambda
\bigg] = 0.\label{10D_Gaugino_Eq}
\end{eqnarray}
We are interested in the solution of these field equations.
Since the action of Eq.(\ref{S_Hetero}) is invariant under the supersymmetry transformation, the fermionic variations of Eqs.(\ref{GrV}), (\ref{GaV}) and (\ref{DlV}) are possible candidates for fermionic zeromodes.
As we have seen in the previous section, the heterotic fivebrane holds a half of supersymmetry by $\epsilon_+$.
Therefore, the supersymmetry variations by $\epsilon_-$ are relevant to the fermionic zeromodes.
The explicit formulae of these variations in the heterotic fivebrane backgrounds are
\begin{eqnarray} %%% Fermionic Zeromodes %%%
\psi_\rho &=& 
(\partial_\rho + \partial^\sigma \Phi{\gamma_\rho}_\sigma) \eta, \label{Gravitino_Var} \\
\lambda &=& 
-\frac{1}{2} \partial_\mu \Phi \gamma^\mu \eta, \label{Dilatino_Var} \\
\chi^I &=&
 -\frac{1}{4} {F^I}_{\mu\nu} \gamma^{\mu\nu} \eta, \label{Gaugino_Var}
\end{eqnarray}
where $\eta$ is an arbitrary Weyl spinor with negative four-dimensional chirality, $I$ is the $SU(2)$ adjoint index, and ten-dimensional gamma matrix $\Gamma$ is replaced by the four-dimensional one $\gamma$.
The field equations for fermionic zeromodes can be obtained by simply neglecting their dependence along the fivebrane.
This simple dimensional reduction of Eqs.(\ref{10D_Gravitino_Eq}), (\ref{10D_Dilatino_Eq}) and (\ref{10D_Gaugino_Eq}) to four-dimensional space gives the following equations.
\begin{eqnarray}  %%% 4 dimensional field equations %%%
%%% gravitino part %%%
\gamma^{\mu\nu\rho} \left(D_\nu - \frac{1}{2} \partial_\nu \Phi \right) \psi^-_\rho
+2\partial^{[\mu} \Phi \gamma^{\nu]} \psi^-_\nu
+4\gamma^{\mu\nu} (D_\nu - \partial_\nu) \lambda_+
-6\partial^\mu \Phi \lambda_+ = 0, \label{Gravitino_Eq} \\
%%% gaugino part %%%
\gamma^\mu \left( {\cal D}_\mu - \partial_\mu \Phi \right) \chi_-^I
+\frac{1}{4} \gamma^\mu \gamma^{\nu\rho} F^I_{\nu\rho} \psi^-_\mu
+\frac{1}{6} \gamma^\mu \gamma^{\nu\rho} F^I_{\nu\rho} \gamma_\mu \lambda_+ = 0, \label{Gaugino_Eq} \\
%%% dilatino part %%%
-8\gamma^\mu \left( D_\mu - \frac{3}{2} \partial_\mu \Phi \right) \lambda_-
-2\gamma^{\mu\nu} (D_\mu + \partial_\mu \Phi) \psi^-_\nu
-\partial^\mu \Phi \psi_\mu
-\frac{1}{4} F_{\mu\nu} \gamma^{\mu\nu} \chi_-^I = 0. \label{Dilatino_Eq}
\end{eqnarray}

\section{Explicit Calculations}
\label{sec:four}
In this section, we find the explicit form of the solution of the fermionic field equations and check its validity as physical fermionic zeromodes.

First, we solve the dilatino field equation of Eq.(\ref{Dilatino_Eq}).
It contains a term $-\frac{1}{4} F^I_{\mu\nu} \gamma^{\mu\nu } \chi^I$ whose order in the derivative expansion does not coincide with the one of the other terms.
We can neglect this higher-derivative term, since all the fields are considered to be leading order in the derivative expansion of the full heterotic supergravitiy theory.
Therefore, the dilatino field equation which we should solve is
\begin{eqnarray}  %%% Reduced Dilatino field equation %%%
-8\gamma^\mu \left( D_\mu - \frac{3}{2} \partial_\mu \Phi \right) \lambda
-2\gamma^{\mu\nu} (D_\mu + \partial_\mu \Phi) \psi_\nu
-\partial^\mu \Phi \psi_\mu = 0. \label{Reduced_Dilatino_Eq}
\end{eqnarray}
It is easy to see that supersymmetry variation of Eqs.(\ref{Gravitino_Var}), (\ref{Dilatino_Var}) and (\ref{Gaugino_Var}) are the solution of the field equations of  (\ref{Gravitino_Eq}), (\ref{Gaugino_Eq}) and (\ref{Reduced_Dilatino_Eq}).

Notice that the kinetic terms of dilatino and gravitino fields are not diagonalized in the action of Eq.(\ref{S_Hetero}) and the four-dimensional field equations of Eqs.(\ref{Gravitino_Eq}) and (\ref{Reduced_Dilatino_Eq}).
We can diagonalize this kinetic mixing in Eqs.(\ref{Gravitino_Eq}) and (\ref{Reduced_Dilatino_Eq}) with the field redefinition of
\begin{eqnarray} %%% Diagonalization of Kinetic Mixing %%%
\psi_\mu \to \psi_\mu - 2\gamma_\mu\lambda.
\end{eqnarray}
The solution of the diagonalized field equations is 
\begin{eqnarray}
\psi_\mu &=& (\partial_\mu-\partial_\mu\Phi)\eta, \\
\lambda  &=& -\frac{1}{2}\partial_\mu\Phi\gamma^\mu\eta, \\
\chi^I     &=& -\frac{1}{4}F^I_{\rho\sigma}\gamma^{\rho\sigma}\eta.
\end{eqnarray}

The explicit form of the spinor $\eta$ has not been determined yet.
It can be fixed by imposing the gauge condition to the gravitino solution.
$\gamma$-tracelessness, $\gamma^\mu\psi_\mu=0$, is usually used as a gauge condition, but in the present case the other gauge condition, namely $D^\mu\psi_\mu=0$, gives simpler result.
This gauge condition requires $\eta$ to be a constant $\eta_0$, and the explicit form of the solution is
\begin{eqnarray}   %%% Fermionic Zeromodes %%%
\psi_\mu &=& -\partial_\mu \Phi \eta_0, \\
\lambda  &=& -\frac{1}{2} \partial_\mu \Phi \gamma^\mu \eta_0, \\
\chi       &=& -\frac{1}{4} F^I_{\rho\sigma} \gamma^{\rho\sigma} \eta_0.
\end{eqnarray}
The normalizability of this solution in the four-dimensional space can be checked easily.
We can find two zeromodes, since $\eta_0$ is defined as a four-dimensional Weyl spinor.
Here, notice that these zeromodes are derived as supersymmetry variations.
Therefore, there is a possibility that the zeromodes are gauged away and unphysical.
But as we will see in the followings, they cannot be gauged away.
The gauge parameter which define a gauge transformation should be normalizable.
In the present case, corresponding gauge parameter $\eta_0$ is not normalizable, since it is a constant.
Therefore, $\eta_0$ does not define gauge transformation.
This means that the zeromodes can not be gauged away, and they are physical.
We can expect some fermion pair condensations since the number of the obtained zeromodes is exactly two.

\section{Conclusion}
\label{sec:five}
We found two physical fermionic zeromodes in the heterotic fivebrane backgrounds by explicitly solving the fermionic field equations.
Two gravitino zeromodes are found in contrast with the result in Ref.\cite{Bellisai}, in which no gravitino zeromodes were found using the extended index formula.
A possible scenario of the gravitino condensation is as follows.

It has been pointed out in Ref.\cite{GSW} that a four-fermion term, which appears in the next order of the $\alpha'$ expansion in the action of the ten-dimensional heterotic supergravity, can give a perfect square term
\begin{eqnarray}
(H_{MNP} - \frac{g^2\phi}{12}tr(\overline{\chi}{\Gamma}_{MNP}\chi))^2. \label{PSQ}
\end{eqnarray}
If we can demand that the cosmological constant vanishes in the heterotic fivebrane backgrounds, the gaugino bilinear $tr (\overline{\chi} {\Gamma}_{MNP} \chi)$ in Eq.(\ref{PSQ}) should have the vacuum expectation value to compensate the non-vanishing field strength $H_{MNP}$.
On the other hand, the supersymmetry transformations of the gravitino and dilatino have the following next order corrections in the $\alpha'$ expansion\cite{BdR}.
\begin{eqnarray}
\delta\psi_M \sim 
\frac{1}{192} \Gamma^{NPS} \Gamma_M \epsilon \left[ tr(\overline{\chi} \Gamma_{NPS} \chi)
+\overline{\psi}^{ab} \Gamma_{NPS} \psi_{ab} \right], \label{NGr}\\
\delta\lambda \sim 
\frac{1}{384} \sqrt{2} \Gamma^{NPS} \epsilon \left[tr(\overline{\chi} \Gamma_{NPS}\chi) 
+\overline{\psi}^{ab} \Gamma_{NPS} \psi_{ab}\right]. \label{NDl}
\end{eqnarray}
From this transformation rule, it is clear that the gaugino condensation $tr \langle \overline{\chi} {\Gamma}^{MNP} \chi \rangle \neq 0$ breaks all supersymmetry.
But the heterotic fivebrane should be invariant under a half of the supersymmetry transformations induced by $\epsilon_+ \in (4, 2_+)$.
Therefore, the gravitino should condensate so that the gaugino condensation are cancelled out in Eqs.(\ref{NGr}) and (\ref{NDl}).
The fact that the existence of both gravitino and gaugino zeromodes could suggest such a phenomenon.

\section{acknowledgments}
The authors would like to thank S. Saito for fruitful discussions and comments.
N.K. would like to thank S.-J.Rey for useful comments.

\end{document}